\documentclass{article}
\usepackage[utf8]{inputenc}
\usepackage{authblk}
\usepackage{setspace}
\usepackage[margin=1in]{geometry}
\usepackage{graphicx}
\graphicspath{{./figures/}}
\usepackage{subcaption}
\usepackage{amsmath}

\usepackage[super,sort&compress,comma]{natbib}
\title{Temperature switchable self-propulsion activity of liquid crystalline microdroplets}

\author[1*]{Manoj Kumar}
\author[1]{Siddharth Sane}
\author[1]{Aniruddh Murali}
\author[1,2]{Shashi Thutupalli}

\affil[1]{Simons Centre for the Study of Living Machines, National Centre for Biological Sciences, Tata Institute of Fundamental Research, GKVK Campus, Bellary Road, Bangalore 560065.}
\affil[2]{International Centre for Theoretical Sciences, Tata Institute of Fundamental Research, Hoskote Village, Bangalore.}
\affil[*]{Corresponding author: manojk@ncbs.res.in}

\date{}

\begin{document}
{
\maketitle

\begin{abstract}
We report on a switchable emulsion droplet microswimmer by utilizing a temperature-dependent transition of the droplet phase. The droplets, made from a liquid crystalline (LC) smectic phase material ($T =$~25~$^{\circ}$C), self-propel only in their nematic and isotropic phases at elevated temperatures ($T\ge$~33.5~$^{\circ}$C). This transition between motile and non-motile states is fully reversible - in the motile state, the droplets exhibit persistent motion and directional memory over multiple heating-cooling cycles.
Further, we distinguish the state of rest from the state of motion by characterizing the chemical and hydrodynamic fields of the droplets. Next, we map the motility behaviour of the droplets across varying surfactant concentrations and temperatures, observing that swimming occurs only at sufficiently high surfactant concentrations above and temperatures above the smectic-nematic phase transition temperature \textit{i.e.} $T\ge$~33.5~$^{\circ}$C. 
Our work envisions the potential of LC emulsion droplets as switchable microswimmers.  
\end{abstract}

\section{Introduction}
In recent years, synthetic microswimmers have garnered significant attention due to their ability to self-propel in fluid environments at low Reynolds numbers (low Reynolds number, $\Re < 1$) without the need for complex biological machinery, such as that employed by flagellated microorganisms\cite{marchetti2013hydrodynamics,shields2017evolution}. Artificial microswimmers have demonstrated the capacity to emulate the complex behaviour of biological swimmers in response to changes in their environments such as multigate or multimodal activity \cite{hokmabad2021emergence}, chemotaxis\cite{jin2017chemotaxis}, electrotaxis\cite{buness2024electrotaxis}, magnetotaxis \cite{wagner2024magnetotaxis} and oscillatory rheotaxis\cite{bretherton1961rheotaxis,dey2022oscillatory}. There are a plethora of artificial microswimmers have been discovered to harvest energy from their surroundings and dissipate it to autonomously fuel their motion\cite{shields2017evolution}.
Several phoretic mechanisms have been reported by which an active particle can self-propel\cite{shields2017evolution,michelin2023self}, but our particular interest is on an active droplet system in which droplet propulsion is triggered by spontaneously generated Marangoni stresses. These stresses arise from surface tension gradients set along the droplet interface, typically caused by chemical reactions at the droplet interface\cite{thutupalli2011swarming, thutupalli2013towards} or by solubilization of the droplets\cite{peddireddy2012solubilization, peddireddy2014liquid, herminghaus2014interfacial,michelin2023self}. Such droplet microswimmers can be created using simple experimental techniques, such as emulsification to produce either water-in-oil or oil-in-water droplets, immersed in a surfactant-containing oil or aqueous medium.\cite{thutupalli2011swarming,izri2014self,michelin2023self}. 

Such artificial microswimmers are gaining prominence, not only for the role they play in fundamental understanding of non-equilibrium dynamics but also for their potential applications\cite{weibel2005microoxen,zhang2010controlled,masoud2012designing}. The development of smart microswimmers opens pathways to future technologies capable of autonomously performing a range of complex tasks, such as targeted drug delivery, microsurgery, surface cleaning agents, biosensing, and micromixing\cite{mhanna2014artificial,peyer2013magnetic,bunea2020recent,yan2022soft}. Consequently, the fabrication and discovery of microswimmers that can respond to specific stimuli—whether chemical, magnetic, thermal, or pH-based—are of great importance.  However, designing self-propelled, autonomously active droplets with controllable motion or the ability to regulate their activity on demand remains a significant challenge. Nevertheless, limited experimental approaches have been reported to investigate the response of microswimmers to external stimuli, such as thermal responses\cite{yoshida2020soft,tu2017self,ramesh2025frozen,cholakova2021rechargeable}, light responses\cite{kaneko2017phototactic,lancia2019reorientation}, and chemical responses \cite{jin2017chemotaxis,sridhar2022light}. In recent years, several switchable active systems have been developed, demonstrating the ability to reversibly control the direction of motion or propulsion trajectories through phase transitions induced by external stimuli, such as light or electric fields\cite{soma2020phase, vutukuri2020light, chen2020situ, al2022guide}.
Controlling the propulsion dynamics of the droplet microswimmers, however, remains difficult without employing external agents to deplete their energy reservoirs. Recently, it was demonstrated that droplet self-propulsion could be reversibly arrested using a temperature-sensitive co-surfactant solution\cite{ramesh2025frozen}. In this study, we report temperature-controlled activity of emulsion droplet microswimmers. We employed thermotropic liquid crystalline oil droplets immersed in a supramicellar surfactant solution, where the droplets remained inactive at ambient temperatures but became motile above a critical temperature (at the phase transition temperature). This transition is thermoreversible, allowing a precise control on droplet motility. The swimming droplets exhibit a strong memory of their previous direction of motion, displaying persistent motion and smooth trajectories.

\section{Results}

\subsection{Self-propelling emulsion droplets due to thermally induced droplet phase transition}

In our experiments, 8CB emulsion droplets prepared in ~0.25~wt\% ionic surfactant solution remained stable at room temperature ($T =$~25~$^{\circ}$C), as shown by the bright field and cross-polarization microscopy images, see Fig.\ref{Figure1}(A-C). 
\begin{figure}[h]
\centering
  \includegraphics[height=6cm]{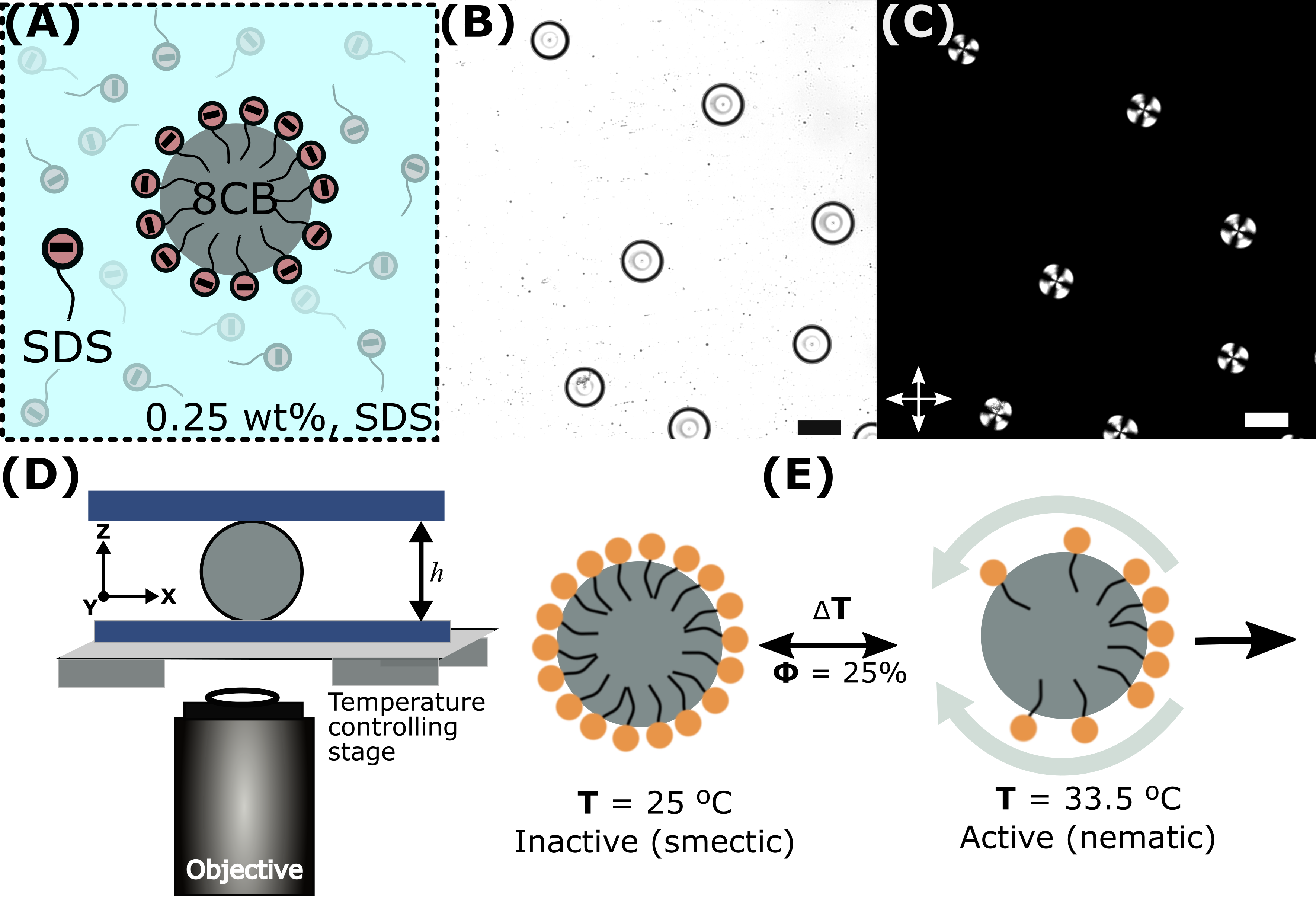}
  \caption{ \textbf{Formation of 8CB emulsion droplets and experimental set up for thermally powering the droplets}: \textbf({A}) Schematic of 8CB emulsion droplet that remains stable in ~0.25~wt\% SDS surfactant solution. \textbf({B}) Bright field microscopy image of 8CB emulsion droplets (scale bar: 50~$\mu m$) and \textbf({C}) the cross-polarization image of droplets. \textbf({D}) Quasi-2D flow cell of height $h\sim d$ ($d$ = droplet diameter) placed under a microscope equipped with a temperature-controlling heating stage. \textbf({E}) Schematics of thermally-induced reversible phase transition (smectic to nematic and vice versa) in 8CB emulsion droplets(droplet size, d = 50~$\mu m$).}
  \label{Figure1}
\end{figure}
These droplets can be stored for several months at room temperature and remain stable. Next, when 8CB droplets were immersed in a supra-micellar surfactant solution ($\phi\sim$ 25~wt\%) of SDS at room temperature ($T =$~25~$^{\circ}$C), unlike 5CB droplets\cite{herminghaus2014interfacial}, the 8CB droplets remained stable and no solubilization was observed (Fig.\ref{Figure1}B). However, when the droplets were heated using a temperature-controlling heating stage under the microscope, the droplets began to self-propel at a temperature $T \ge$~33~$^{\circ}$C (Fig.\ref{Figure1}D and E). The swimming droplets showed a thermoreversible nature, as the droplets start swimming only after reaching a critical temperature ($T \approx$~33.5~$^{\circ}$C) and stop swimming as soon as the temperature is maintained below that critical temperature, see supplementary videos SV1-SV2. It is important to note that the 8CB droplets exhibit this characteristic behaviour only with a supramicellar solution of ionic surfactants. In a non-ionic surfactant, viz. 5~wt\% Triton X-100, the 8CB droplet shows temperature-independent swimming with chaotic motion at room temperature. Thus, it suggests a scope for further study on the role of non-ionic surfactants in controlling the 8CB droplet solubilization, see supplementary video SV3).\\

The self-propulsion mechanism of 8CB emulsion droplets in a supramicellar surfactant solution is similar to that reported for other oil droplets systems, such as 5CB in SDS or 5CB in TTAB surfactant solution \cite{thutupalli2018flow,dey2022oscillatory}, except for the thermal energy required for the phase transitions in the 8CB droplet. Briefly, the onset of droplet swimming in a supramicellar surfactant solution ($\phi\sim$ 25~wt\%) at $T\approx$~33.5~$^{\circ}$C occurs due to the migration of oil molecules into SDS micelles, thus initiating the dissolution of 8CB droplets. This results in the generation of filled micelles in the droplet vicinity and increases the interfacial tension along the droplet interface. The advection perturbations generate a gradient in surface tension along the droplet interface because of the non-uniform surfactant coverage. A self-sustained gradient caused by the Marangoni flows generated along the droplet interface makes the droplet move forward, leaving behind a persistent trail of filled micelles. The solubilisation rate of the 8CB droplet in 25~wt\% SDS at $T\approx$~33.5~$^{\circ}$C is $\approx$~3~$nm/s$). This corresponds to a droplet size (d) decrease of $\approx$ 2~$\mu m$ over a period of $\approx$ 10 minutes. Notably, this rate of solubilization is expected to further increase with rising temperatures, as 8CB solubilization dynamics are strongly temperature-dependent.\\

\begin{figure*}[h!]
\centering
  \includegraphics[height=5cm]{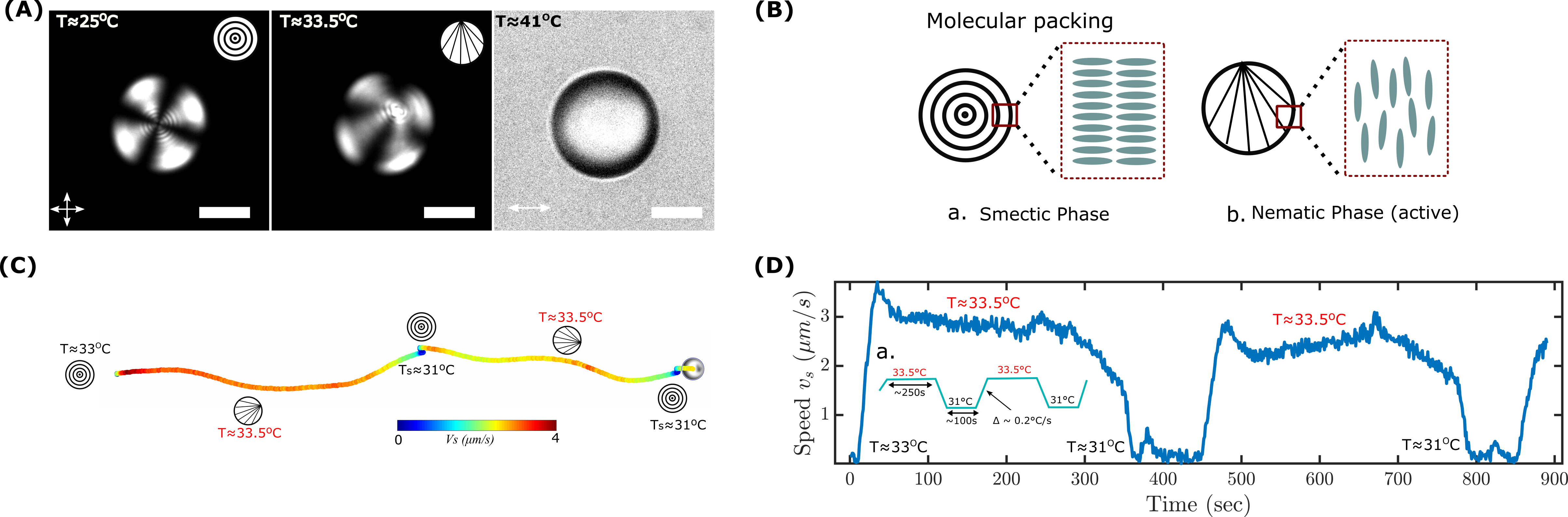}
  \caption{\textbf({A}) Focal-conic or fan-like texture of 8CB emulsion droplet represents smectic phase under cross-polarisation (\textit{left panel}) microscopy. At temperature, $T=$~25~$^{\circ}$C droplet is non-motile (\textit{left panel}), the droplet is in nematic phase at $T\approx$~33.5~$^{\circ}$C self-propels with a defect pointing in the propulsion direction (\textit{middle panel}), and at $T\approx$~41~$^{\circ}$C self-propelling droplet is in isotropic phase (scale bar = 25~$\mu m$). (\textbf{B}) a. Molecular packing (8CB molecules) shows highly ordered smectic (non-active) phase at $T=$~25~$^{\circ}$C, b. less ordered nematic phase (active) with defect orientation in the propulsion direction at $T\approx$~33.5~$^{\circ}$C. (\textbf{C}) The trajectory shows the straight and temperature-reversible motion of the swimming 8CB droplets (droplet size, d = 80~$\mu m$, confinement height in Hele-shaw cell is h $\approx$ 80~$\mu m$).(\textbf{D}) Tuning of droplet propulsion speed with temperature heating-cooling cycle. Inset (\textbf{a.}) in (\textbf{D}), the curve shows the heating/cooling cycle and the corresponding heating, cooling and holding time where the rate of heating/cooling is $\Delta$~$\approx$~0.2~$^{\circ}$C/s.}
  \label{Figure2}
\end{figure*}

The 8CB emulsion droplets, which are in their smectic phase at room temperature ($T =$~25~$^{\circ}$C) when immersed in a surfactant solution $\phi\sim$ 25~wt\% remained non-motile. Since 8CB LC is thermotropic, the droplet phase transitions from smectic to nematic at $T\approx$~33.5~$^{\circ}$C and isotropic phase at $T\approx$~41~$^{\circ}$C. Surprisingly, 8CB droplets start swimming in their nematic phase ($T\approx$~33.5~$^{\circ}$C) and continue to swim even in their isotropic phase, see supplementary videos SV1-SV2 and Fig.\ref{Figure2}A. As we lower the temperature below $T\approx$~33.5~$^{\circ}$C, the droplets stop swimming as they transition back to their smectic phase. This shows that 8CB oil droplets can swim only in their nematic and isotropic phases resulting upon heating of the smectic phase (at room temperature). The droplet swimming behaviour also displays the thermoreversible nature, where the temperature acts like a switch to stop and resume the swimming droplets when required. Next, we recorded and analyzed the trajectories of swimming 8CB droplets over multiple heating and cooling cycles as the droplet transitioned between the non-swimming smectic and swimming nematic phases, see Fig.\ref{Figure2}C-D and supplementary video SV4. Unlike nematic 5CB oil droplets ($T =$~25~$^{\circ}$C), which exhibit curling trajectories and ballistic motion scaling as  $\propto$ $t^2$ over shorter time scale, as previously reported\cite{kruger2016curling}, the 8CB emulsion droplets in their nematic phase ($T\approx$~33.5~$^{\circ}$C) display smooth, non-curling trajectories with persistent motion(Fig.\ref{Figure2}C). Although both the 5CB and the 8CB emulsion droplets self-propel in their nematic phase, the 8CB droplets do not show curling instabilities despite nematic ordering\cite{kruger2016curling}. The straight or non-curling motion is achieved when the droplet is in the isotropic phase or has a slower propulsion speed in the nematic phase. The 8CB droplets in their nematic phase the self-propulsion speed is slower ($v_s\approx$~3~$\mu m$/s) compared to the 5CB droplets ($v_s\approx$~8~$\mu m$/s)\cite{kumar2024emergent} resulting in smooth and straight trajectories. This swimming speed is significantly lower compared to the previously reported droplet systems of a similar size (d = 80~$\mu m$)\cite{bechinger2016active}. This reduced speed is attributed to the lower solubilization rate of 8CB droplets in SDS ($\approx$ 3~nm/s), the propulsion speed depends upon the rate of solubilisation\cite{peddireddy2014liquid}.
When the 8CB emulsion droplets were subjected to temperature-controlled start-stop cycles, they consistently followed the same straight trajectory, indicating that the droplets retained a memory of their previous positions and directions of motion (Fig.\ref{Figure2}C and supplementary video SV4). To rationalize this memory effect, we extended the stagnation time before reinitiating the droplet motion. Remarkably, even after a stagnation period of up to ~30~minute, the droplets continued to swim in the same direction as before. Beyond ~30~minute of stagnation, we observed a net drift in the sample, and thus, experiments were not extended beyond this time frame (supplementary video SV5). The memory effect is observed exclusively in the nematic phase during the transition between the smectic and nematic phases. However, this effect is absent in the isotropic phase, where transitions occur among the isotropic, nematic, and smectic phases (supplementary video SV6).

Next, we measured the speed of the 8CB emulsion droplet during the run-and-stop thermal cycle. A sudden increase in speed was observed when the temperature reached $T\approx$~33.5~$^{\circ}$C, with the droplets propelling at a speed of $v_s\approx$~3~$\mu m$/s, which decays to $v_s\approx$~0 as the temperature was lowered to $T\approx$ 31~$^{\circ}$C (below $T_c\approx$ 33.5~$^{\circ}$C), see Fig.\ref{Figure2}D and inset a. and supplementary video SV4. Interestingly, similar dynamics of sharp increase in speed followed by a short decay is observed in recent experiments that employ a different mechanism\cite{ramesh2025frozen} which suggests that the underlying hydrodynamic mechanisms could be very similar. After a waiting period of $t\approx$ 100~s (during which the droplet remained at rest as long as the temperature was unchanged), increasing the temperature back to $T\approx$~33.5~$^{\circ}$C caused another rapid jump in speed from $v_s\approx$~0 to $v_s\approx$~3~$\mu m$/s, and the droplet resumed swimming at a steady speed until the temperature was decreased again. Slight fluctuations in droplet speed were noted when the droplet was stagnant, likely due to some drift present in the sample. 

\subsection{Chemical and flow fields}

We now investigate the dynamics of droplet self-propulsion by examining the chemical and hydrodynamic fields. In a typical self-propelling droplet system, droplet dissolution and propulsion produce a chemical field trail and fluid flow around the droplet. The chemical field is the result of spent fuel, which generates oil-filled surfactant micelles by transferring oil molecules from the droplet phase into empty micelles present in the surrounding aqueous medium.
\begin{figure}[h]
\centering
  \includegraphics[height=8cm]{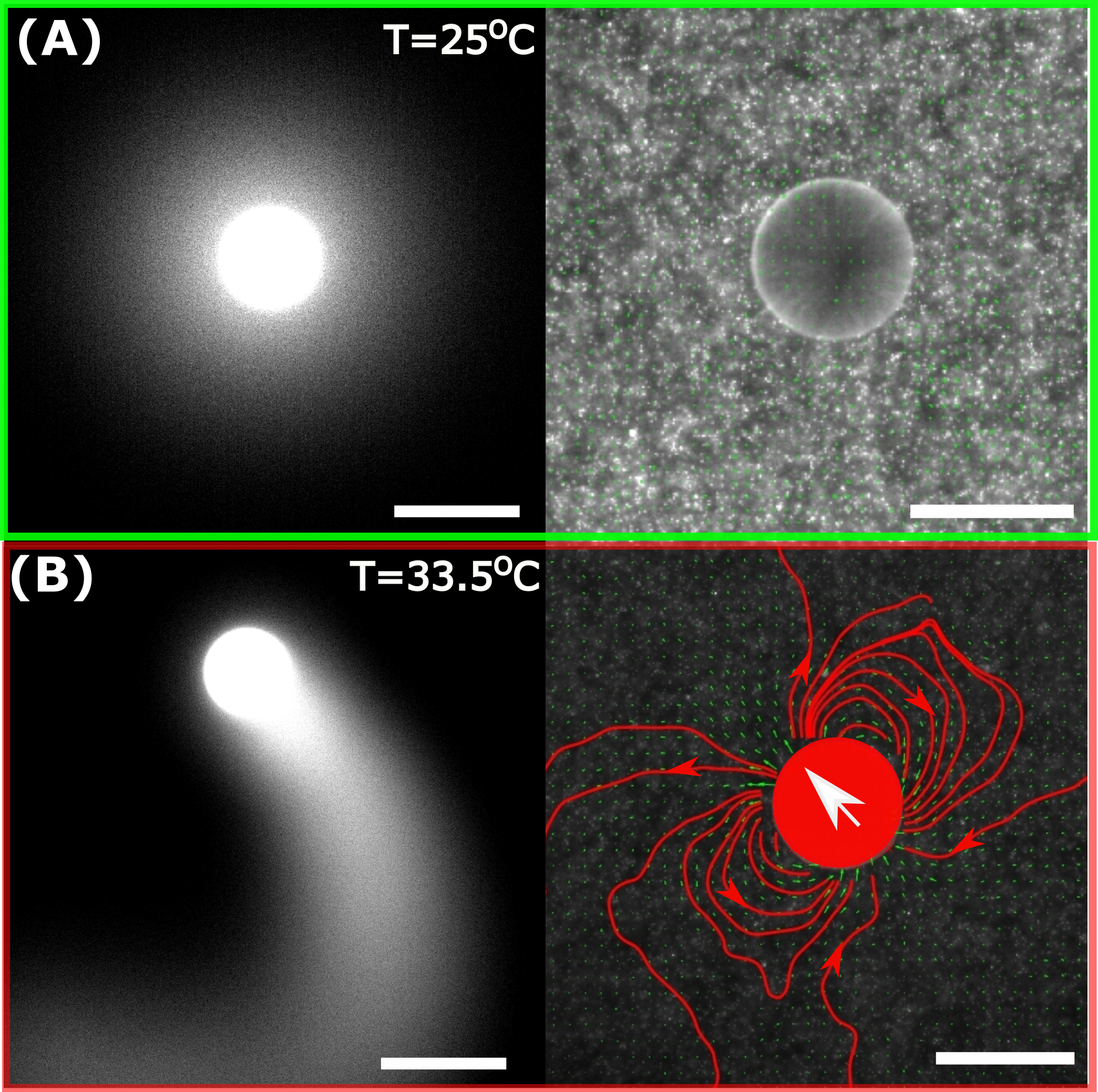}
  \caption{\textbf({A}) Top panel shows the chemical field (\textit{left panel}) and hydrodynamic field (\textit{right panel}) of 8CB emulsion droplet at $T =$~25~$^{\circ}$C and surfactant concentration $\phi=$~25~wt\% (scale bar: 50~$\mu m$). \textbf({B}) shows the chemical field (\textit{left panel}) and hydrodynamic field (\textit{right panel}) of 8CB emulsion droplet at $T = $~33.5~$^{\circ}$C and surfactant concentration $\phi=$ 25~wt\% (scale bar: 50~$\mu m$). The droplet size, d = 50~$\mu m$ and confinement height, h $\approx$ 50~$\mu m$.}
  \label{Figure3}
\end{figure}

Chemical fields were visualized using an oil-soluble fluorescent dye, Nile red, that was premixed with the oil phase. No observable chemical field was produced by the droplet when in its smectic phase ($T =$~25~$^{\circ}$C), the uniform fluorescence signal from the droplet suggesting that the dye mostly remained contained within the droplet and only slightly diffused into the aqueous phase outside (Fig.\ref{Figure3}A, \textit{top, left panel}, supplementary video SV7). A curve in the chemical field trail can be seen for the swimming droplet in Fig. \ref{Figure3}B. This behaviour arises due to the presence of neighbouring droplets in the initial region, causing deflections in the trajectory. However, as the droplet enters sparser regions or moves away from crowded regions, it exhibits a straight and smooth trajectory. Next, the size of SDS micelles is estimated after measuring the diffusivity of the oil-filled micelles. The $\mathcal R = (k_BT/ 6\pi\eta D)$ $\approx$ 2.4~nm is the radius of the oil-filled micelles, where D $\approx$ 18~$\mu m^2$/s, $\eta$ = 5.2~mPa.s, being the viscosity of the solvent, and temperature $T = $~33.5~$^{\circ}$C. The hydrodynamic fields were also investigated using fluorescent tracer particles, which revealed the absence of flow generation in the smectic phase (Fig.\ref{Figure3}A, \textit{top, right panel}, supplementary video SV8). In contrast, a prominent trail of the chemical field in the wake of the droplet, along with a dipolar flow field, confirmed that 8CB droplets exhibit motility in the nematic phase ($T\ge$~33.5~$^{\circ}$C) (Fig.\ref{Figure3}B, \textit{bottom panel}, supplementary video SV7). Thus, our observations show that the onset of dissolution processes in the nematic phase appears to drive droplet dynamics.

\begin{figure*}[ht!]
\centering
  \includegraphics[height=5cm]{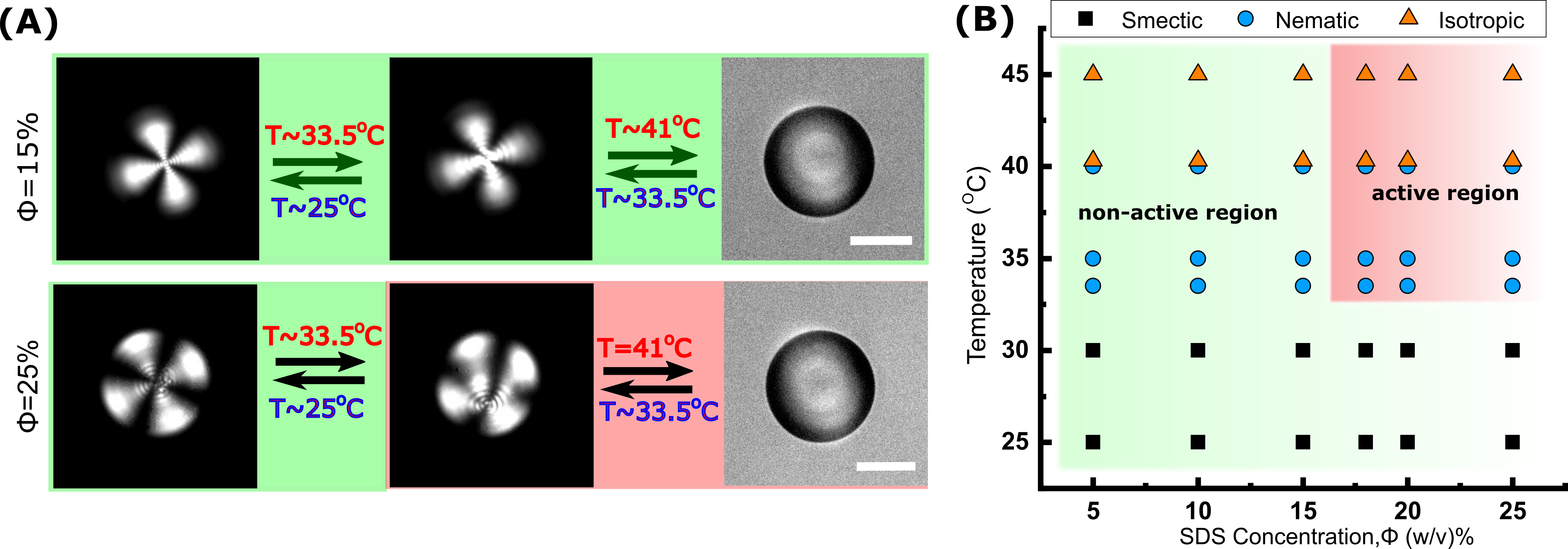}
  \caption{\textbf({A}) The cross-polarised microscopy images of 8CB emulsion droplets at $\phi=$~15~wt\% shows thermoreversible nature. The images of droplets at different phase transition temperatures (\textit{top panel, left to right}) $T =$~25~$^{\circ}$C (smectic phase), $T\approx$~33.5~$^{\circ}$C (nematic phase) and $T\approx$~41~$^{\circ}$C (isotropic phase) and vice-versa. At $\phi=$~25~wt\%, the crossed-polariser microscopy images of 8CB emulsion droplets show inactive smectic phase at $T =$~25~$^{\circ}$C, active nematic phase at $T\approx$~41~$^{\circ}$C and active isotropic phase at $T\approx$~41~$^{\circ}$C (scale bar: 25~$\mu m$).\textbf({B}) Phase diagram shows the motile and non-motile regions of 8CB emulsion droplet as a function of temperature ($T$) and surfactant concentration ($\phi$).The droplet size, d = 50~$\mu m$ and confinement height, h $\approx$ 50~$\mu m$.}
  \label{Figure4}
\end{figure*}

\subsection{Mapping of active and non-active regions in a temperature versus surfactant concentration phase diagram}

So far, we showed that 8CB emulsion droplets self-propel at $T\ge$~33.5~$^{\circ}$C (only in their nematic and isotropic phases) when immersed in a $\phi=$~25~wt\% surfactant solution. We now turn to mapping the active and non-active states of the droplets as a function of temperature and surfactant concentration.
 Droplets in the surfactant concentration range of $\phi\ge$~5~wt\% to $\phi\le$~15~wt\% remained inactive across all temperatures, despite the phase transitions from smectic (T = 25~$^{\circ}$C) to nematic ($T\approx$ 33.5~$^{\circ}$C) and from nematic to isotropic ($T\approx$ 41~$^{\circ}$C), see Fig.\ref{Figure4}A (\textit{top panel}) and Fig.\ref{Figure4}B, supplementary video SV1-SV2. However, when droplets were immersed in a surfactant solution with concentration, $\phi\ge$ 18~wt\% and temperature was varied from $T =$~25~$^{\circ}$C to $T =$~41~$^{\circ}$C, 
 the 8CB droplets remained inactive in the smectic phase but became active in the nematic phase ($T\approx$ 33.5~$^{\circ}$C). This defines a narrow region of activity in the 8CB phase diagram, where self-propulsion occurs at high surfactant concentrations ($\phi\ge$~15~wt\%} and elevated temperatures ($T\ge$~33.5~$^{\circ}$C). Therefore, 8CB droplets in the smectic phase remained inactive at room temperature regardless of surfactant concentration, becoming active only above a critical temperature range (Fig.~ \ref{Figure4}B).

 \subsection{Temperature and size dependency of droplet swimming speed}

\begin{figure*}[h!]
\centering
  \includegraphics[height=7cm]{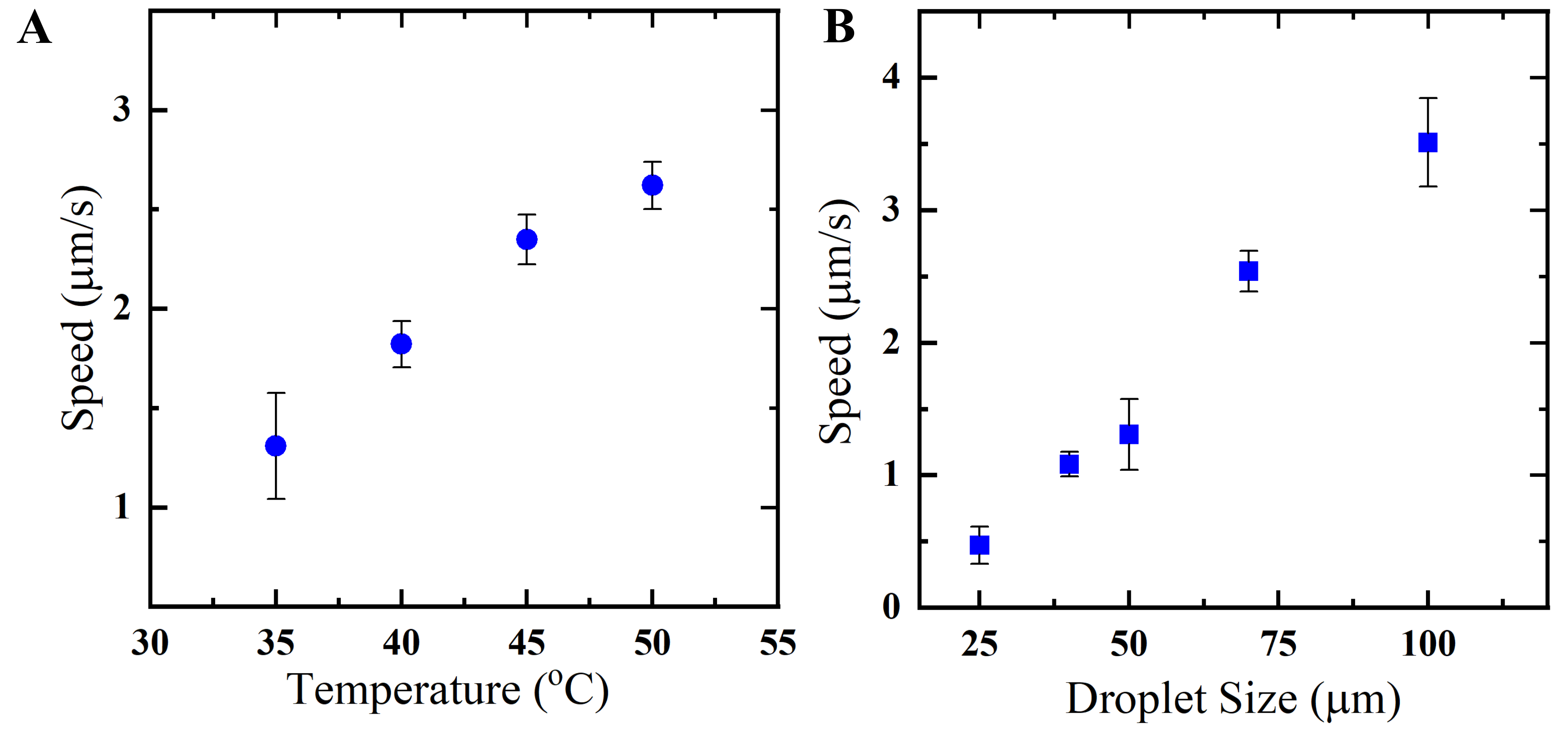}
  \caption{\textbf({A) The swimming speed of 8CB droplets was measured as a function of temperature upon heating droplets (d = 50~$\mu m$) in a quasi-2D flow cell with a height of h $\approx$ 50~$\mu m$. As the temperature increased from T = $35^oC$ to $50^oC$, the swimming speed of the droplets exhibited a temperature dependence. \textbf({B}}) Plot shows the droplet swimming speed with droplet size (d = diameter) measured in a quasi-2D flow cell. The droplet sizes range from 25~$\mu m$ to 100~$\mu m$.}
  \label{Figure5}
\end{figure*}

 As we notice from the phase diagram shown in Fig.\ref{Figure4}B, the 8CB droplets exhibit motility only in their nematic and isotropic phases, which emerges at elevated temperatures. To further investigate, we measured the effect of temperature on the swimming speed of the droplets. The swimming speed of the droplets increases with temperature, as shown when the temperature was raised from $T = $~35~$^{\circ}$C to ~50~$^{\circ}$C, see Fig.\ref{Figure5}A. These results indicate that the swimming speed of the droplets not only increases during the transition from the nematic to isotropic phase but also within the nematic or isotropic phase. This behavior underscores the strong temperature dependence of droplet motility. The swimming speed of the droplets also exhibits a dependence on their size. Droplets of different sizes were prepared via sonication, with diameters ranging from 25~$\mu m$ to 100~$\mu m$. The smallest droplets (d = 25~$\mu m$) displayed a motility speed of $v_s\approx$~0.47~$\rm \pm {0.14}~{\mu m/s}$, while larger droplets (d = 100~$\mu m$) exhibited an increased swimming speed of $v_s\approx$~3.5~$\rm \pm {0.33}~{\mu m/s}$. Our results reveal a linear relationship between swimming speed and droplet size, as shown in Fig.\ref{Figure5}B.

Finally, we summarize the underlying reason for the non-swimming nature of the 8CB droplets in their smectic phase and the plausible reason for memory. In the smectic phase, 8CB molecules exhibit a highly ordered molecular arrangement due to the presence of both positional and orientational order (Fig.\ref{Figure2}B), which significantly limits the solubility of the smectic phase in the ionic surfactant solution. However, upon heating, the droplet transitions to the nematic phase—a less ordered state—which initiates solubilization. The swimming speed of the droplet increases with temperature (Fig.\ref{Figure5}A), suggesting that solubilization is temperature-dependent. The swimming speed reaches its maximum in the isotropic phase, where the molecules are in the most disordered state, corresponding to the highest solubility. Consequently, the smectic phase displays no solubilization due to its highly ordered molecular arrangement. Interestingly, with non-ionic surfactants, it may be possible to disrupt the molecular arrangement in the smectic phase, as suggested by our preliminary observations (Supplementary Video SV3). This indicates potential avenues for further exploration into the solubilization mechanisms of smectic phases under different surfactant conditions. Further, the memory displayed by the swimming 8CB droplets during their phase transitions from  smectic to nematic phase could be a result of the residual chemical field. We hypothesize that the residual chemical field serves as a signature of the direction and memory of motion. To investigate further, we conducted experiments aimed at capturing the gradient in the chemical field during the droplet stagnation period. Due to limitations in the resolution of the chemical field measurements, we were unable to detect this gradient conclusively. In addition to the chemical field, the memory effect appears to be influenced by both the size and speed of the droplets, which warrants further exploration to fully understand the underlying mechanisms.

Additionally, we also attempted to investigate the possibility of 8CB droplets swimming in an unbounded fluid medium upon heating the medium. Our preliminary observations show that the droplets can swim in a 3D fluid medium, however, the convective currents generated during the heating introduced perturbations in the droplet swimming dynamics, see the supplementary video SV9. These currents made it challenging to conclusively determine whether the droplets were moving away from the temperature gradient or toward it. Further optimization of the experimental setup to minimize convective effects is required for a more precise assessment.

\section{Conclusions}
We have demonstrated that the dynamics of swimming 8CB droplets can be thermoreversibly controlled by thermally induced phase transitions. The 8CB droplets, when immersed in a supramicellar surfactant solution, remain non-motile in their smectic phase at room temperature. However, the phase transition to the nematic phase initiates droplet solubilization and spontaneous self-propulsion. The non-motile behavior of 8CB droplets in their smectic phase is likely due to the highly ordered molecular packing (exhibiting both positional and orientational order), which inhibits solubilization. As the temperature increases, this order is disrupted, leading to the less ordered nematic phase that allows for solubilization and subsequent propulsion.

The thermotropic nature of 8CB LC oil droplets enables the control of droplet swimming on demand. In addition to persistent motion, the swimming droplets exhibit a strong memory effect, retaining their direction of motion even after being stopped and restarted. This trajectory memory may result from slow propulsion speeds and chemical interactions with the self-generated chemical field, which limit the droplets to straight trajectories — this warrants further investigation.
Our findings also suggest the possibility of exploring other liquid crystalline systems, such as Smectic C, Columnar, and crystalline phases, to understand the influence of molecular order and packing on thermally controlled solubility-driven self-propulsion. The fabrication of stimuli-responsive microswimmers is of both fundamental and applied interest, with thermoresponsive microswimmers offering promise for applications as well as broader biomedical and mechanical applications.

\section{Experimental Section}

\subsection{Materials and Methods}

  4'-Octyl-4-biphenylcarbonitrile (8CB) liquid crystal (LC) and sodium dodecyl sulfate (SDS) were purchased from Sigma-Aldrich. The aqueous solution of SDS was prepared in Milli-Q water of resistivity ~18~$\Omega$cm. To prepare oil-in-water emulsion droplets of 8CB LC oil, we heat 8CB LC oil to ~45~$^{\circ}$C before adding it to the SDS solution, heating makes it easier to pipette. We added ~0.5~g of 8CB oil to ~1~mL of ~0.25~wt\% SDS solution into an ~2~mL eppendorf tube. The emulsion droplets were produced using pipette mixing. For producing uniform-sized emulsion droplets a microfluidic technique was used with a continuous heating of the oil phase under an ITO-heating plate. The stability of the emulsion droplets at room temperature was confirmed using optical microscopy. 

\subsection{Microscopy}

We used brightfield microscopy to visualize 8CB emulsion droplets. Droplets were imaged under an Olympus IX81 microscope with a 4x and 20x UPLSAPO objective and images were captured using a Photometrics Prime camera. The cross-polarized images were recorded by placing the two polarizers, one on the light source and the other beneath the stage. 
For fluorescence images, we used a CoolLED PE-4000 lamp. A 16-bit, fluorescence channel image of size 2048 x 2048 pixels was recorded (excitation: 550~nm, dichroic: Sem-rock’s Quad Band) using a Photometrics Prime camera with a frame rate of 5~fps and exposure time of ~10~ms. For sample-controlled heating, we used an okolab heating stage equipped with a temperature controller (H101-BASIC-BL), temperature range: from ~10 - 15~$^{\circ}$C below room temperature to ~60~$^{\circ}$C with a precision of $\pm$~0.1~$^{\circ}$C. 

\subsection{Flow Cell Fabrication}
We designed a quasi-2D flow cell to confine the 8CB emulsion droplets in 2D and study their dynamics using microscopy. We used a double-sided adhesive tape (3M, USA; product no. 82601) of thickness ~10~$\mu m$ in layer-by-layer fashion between a glass slide and a cover slip to create a flow cell of desired height. The ends of the flow cell were kept sealed during the measurements.\ 

\subsection{Chemical Field and Hydrodynamic field}
To measure the chemical field generated by self-propelling 8CB emulsion droplets, we mixed the oil-soluble dye, Nile red, with 8CB (oil phase). Chemical field images were obtained from videos recorded in the red ($\lambda_{ex}/\lambda_{em}$ = 552/~636~nm) channel of fluorescence microscopy.
Hydrodynamic field was measured by mixing the green fluorescent tracer particles of size ~0.5~$\mu m$ with surfactant solution (2~$\mu L$ of ~5~wt\% tracer particles in ~1~mL of ~25~wt\% SDS solution). The FlowTrace plugin in ImageJ\cite{gilpin2017flowtrace} was used to trace flows, and streamlines were obtained using PIVlab \cite{thielicke2021particle}.\

\subsection{Image analysis}
We recorded the images using bright-field microscopy. Each image is of size $2048\times2048$ pixels. To improve the contrast of the droplets against the background, we generate a background image by averaging all the images in the sequence, which was then subtracted from each frame. After background subtraction, images were processed using custom code in Matlab in the following steps: (1) define a region of interest (ROI) around a droplet and then (2) detect that droplet using the \texttt{Imfindcircles} function in Matlab. To stitch the detected centers in successive frames and generate trajectories, a least-squares tracking algorithm was applied to the current experimental data. Using a conversion factor the trajectories were converted from pixels to micrometers. From these trajectories, the mean squared displacement and the speed of the self-propelling droplets were calculated.

\section*{Acknowledgments}
 We thank Jyoti Prasad Banerjee for discussions. We acknowledge the support from the Department of Atomic Energy (India), under project no.\,RTI4006, the Simons Foundation (Grant No.\,287975), the Human Frontier Science Program and the Max Planck Society through a Max-Planck-Partner-Group. Some of this work was performed as a final year project (S.S.) of the Research Education Advancement Programme (REAP) of the Bangalore Association for Science Education at the Jawaharlal Nehru Planetarium, Bangalore. 

\subsection*{Conflicts of Interest}
There are no conflicts to declare.

\end{document}